\documentstyle[12pt,iopfts]{ioplppt}
\begin{document}
\title{Molecular vibrational states in the binary cold fission 
of $^{252}$Cf}
\author{\c S Mi\c sicu$\dagger$, A S\u andulescu$\dagger\ddagger$ 
and W Greiner$\ddagger$}
\address{$\dagger$ Institute for Nuclear Physics and Engineering, 
Bucharest-M\u agurele, PO Box MG-6, ROMANIA}
\address{$\ddagger$ Institut f\"ur Theoretische Physik der J W Goethe 
 Universit\"at, D-60054, Frankfurt am Main, GERMANY}

\begin{abstract}
We predict a molecular vibrational state in the cold binary fission 
of $^{252}$Cf using a simple decay cluster model. The Hamiltonian of 
two even-even fragments in the pole-pole configuration is built in 
the same fashion as that for the dinuclear molecule formed in heavy-ions 
collisions.
The interaction between the two fragments is described by the 
double-folding M3Y potential. The spectrum of the butterfly vibrations is
derived and its dependence on fragments deformation and mass-assymetry is
discussed. Some experimental implications are commented.   
\end{abstract}

\section{Introduction}

Presently the cold fragmentation of heavy nuclei with masses ranging 
from A$\approx$70 to $\approx$160 atomic units is a subject of 
intensive experimental and theoretical investigations \cite{hkb93} 
\cite{sfc96}. This phenomenon is an extension of the ordinary cluster 
radioactivity to the fission of heavy nuclei. In such processes the emerging 
fragments carry nearly no excitation energy (TXE) so that the kinetic 
energy (TKE) is close to the reaction $Q$-value. 
The binary cold neutronless fission of $^{252}$Cf was recently observed
using the multiple Ge-detector arrays and the $\gamma-\gamma-\gamma$ 
coincidence technique \cite{tak94}. In these experiments the even-even 
fragments were populating the lowest states with $I=0^{+},~2^{+},~4^{+}$
of the $K=0^{+}$ ground-state rotational band, which subsequently decayed 
through $\gamma$ emmision.
The fragmentations involving three final nuclei have been also observed in 
these experiments. The two heavier fragments which have on the average 
20 to 40 MeV of total excitation energy (TXE) are usually accompanyed by a 
light charged particle. This third fragment is most likely to be an $\alpha$ 
particle. The relative transition probabilities to the 2$^{+}$, 
4$^{+}$ and 6$^{+}$ excited states of the ground state rotational band and 
the 5$^{-}$ and 7$^{-}$ states of the octupole band are recorded for even-even 
heavy pairs (Ba-Zr for example). 
Thus the individual fragments resulting from the cold fission process 
display a fine structure of rotational character mainly. 
A natural question then occur : Is it possible that the nascent fragments 
still bound together, close to the scission point,  
form a so called {\em nuclear molecule} similar to those produced by 
colliding heavy ions? The answer is positive if the potential between the 
fragments has {\em pockets} \cite{gps95} as happens for isomeric states.  
On the other hand it is important to remind that the magic radioactivity
consisting of the emission of heavy nuclei, such as $^{14}$C, $^{20}$Ne, 
$^{28,30}$Mg and $^{32}$Si nuclei, which has been predicted in the early 
eightees by S\u andulescu, Poenaru and Greiner \cite{spg80} can be 
portrayed as a particular example of nuclear molecules occuring in the 
fission process.

In the case of the binary cold fission such molecular excitations are more 
likely to occur when the two final nuclei 
are in the touching (nose-to-nose) configuration. The displayed collective 
spectrum should be analogous to that of a nuclear molecule formed by two 
colliding heavy ions sticking together for a short period of time and 
subjected to the interplay of Coulomb repulsion and nuclear attraction.
In addition to the $\beta$- and $\gamma$- vibrational degrees of freedom 
of the individual nuclei, one finds dipole oscillations of the relative 
coordinate \cite{iac81} and rotation-vibration modes of quadrupole nature 
like butterfly- and belly-dancer-type motions.
In the present work we report results only for the butterfly modes (fig.1) 
We obtain their energy spectrum for several even-even splittings of
$^{252}$Cf employing the multipolar form of the M3Y potential for heavy 
deformed nuclei. This allows us to separate the radial from   
the angular variables in the Hamiltonian. Since the radial part 
of the M3Y potential does not contain pockets we disregard at once the 
molecular dipole oscillations. Moreover, since we do not
expect major changes of the ground state deformations of the individual 
fragments during the cold fission process, the $\beta$ and $\gamma$ 
vibrations are excluded to.

\section{The Hamiltonian of the two fragments in the pole-pole configuration}
 
The classical form of the Hamiltonian for two interacting, quadrupole 
deformed nuclei is given by
\begin{equation}
H=T+V(r,\alpha_{2m}^{1},\alpha_{2m}^{2})
\end{equation}
The kinetic energy reads
\begin{equation}
T={1\over 2}B_{1}\sum_{m}\dot\alpha_{2m}^{1}\dot\alpha_{2-m}^{1}+
  {1\over 2}B_{2}\sum_{m}\dot\alpha_{2m}^{2}\dot\alpha_{2-m}^{2}+
  {1\over 2}\mu\dot{\bi r}^{2}
\end{equation}
where the first two terms are giving the kinetic contribution from 
each fragment and the last one accounts for the relative kinetic energy. 
The double-folded deformed potential barrier was expressed in a 
previous paper \cite{sfc96} by performing a general multipole 
expansion of the potential 
\begin{equation}
V({\bi R},\Omega_{1},\Omega_{2}) = \sum_{\lambda_{i},\mu_{i}}
V_{\lambda_{1}\lambda_{2}\lambda_{3}}^{\mu_{1}\mu_{2}\mu_{3}}(R)
D_{\mu_{1}0}^{\lambda_{1}}(\Omega_{1})
D_{\mu_{2}0}^{\lambda_{2}}(\Omega_{2})
Y_{\lambda_{3}\mu_{3}}({\hat R})
\end{equation}
In what follows we will limit ourselves to the case of monopole and 
quadrupole terms $(\lambda_{i}=0,2)$. 
As we mentioned before we consider that both fragments are aligned along 
the same symmetry axis and thus the above equation will have a more simple
form
\begin{equation}
V(R,\phi_{1},\phi_{2})=\sum_{\lambda_{i}=0,2}
V_{\lambda_{1}\lambda_{2}\lambda_{3}}^{0~0~0}(R)
P_{2}(\cos \phi_{1})P_{2}(\cos \phi_{2})
\end{equation}
Because we consider the touching configuration (pole-pole), $\phi_{1}$ 
and $\phi_{2}$ are approximately related through the relation
\begin{equation}
R_{1}\sin\phi_{1} \approx R_{2}\sin\phi_{2}
\end{equation}
where $R_{1}$ and $R_{2}$ are the radii along the symmetry axis 
for prolate nuclei ($R_{i}=R_{0i}(1+\sqrt{5\over \pi}\beta_{0i})$).
For small inclinations angles ($|\phi_{1}|,|\phi_{2}|\ll 1$) and taking 
into account that $\phi_{2}$ has opposite sign as $\phi_{1}$, the above 
equation reads 
\begin{equation}
\phi_{2}\simeq -{R_{2}\over R_{1}}\phi_{1}
\end{equation}
The {\em butterfly} harmonic potential 
\begin{equation}
V_{butt} = {1\over 2}C_{\varepsilon}\varepsilon^2
\end{equation}
can be obtained from the M3Y potential by making a Taylor expansion in
the variable $\varepsilon$ up to the second order, which eventually 
leads us to the expression of the stiffness
\begin{eqnarray}
C_{\varepsilon} & = & -3\left\{ V_{202}(r_{c})+V_{022}(r_{c})
{R_{1}^2\over R_{2}^2}\right.\\
& & + \left.{1\over 2}\left ( 1+{R_{1}^2\over R_{2}^2}\right)
(V_{220}(r_{c})+V_{222}(r_{c})+V_{224}(r_{c}))\right\}
\end{eqnarray}
This quantity is evaluated at the pole-pole configuration where 
$V(r_{c})=V_{butt}$.
It is worthwhile to notice that the monopole part of the potential does 
not contribute to the {\em butterfly} potential. Therefore the 
occurrence of the butterfly potential is strictly connected to 
the deformations of the fragments, i.e. only nuclear molecules with 
elongated shapes are able to experience such oscillations.
It is also important to mention that since at the scission point the 
fragments have compact shapes, we can employ deformations close to the 
ground-state values.
According to \cite{hg84} the zeroth-order Hamilton operator for an 
asymmetric giant molecule reads
\begin{eqnarray}
H'_{0} & = &{\hbar^{2}\over 2\mu r_{0}^{2}}(L^{2}-{L'_{3}}^{2}) +
\frac{\hbar^{2}\left ({L'_{3}}^{2}-{1\over 4}\right )}
{6[B_{1}\beta_{01}^{2}+
{R_{1}^{2}\over R_{2}^2}B_{2}\beta_{02}^{2}]\varepsilon^{2}}\\
& - &\frac{\hbar^{2}}{6[B_{1}\beta_{01}^{2}+
{R_{1}^{2}\over R_{2}^2}B_{2}\beta_{02}^{2}]}
{\partial^{2}\over \partial\varepsilon^{2}}+
\frac{C_{\varepsilon}}{2}\varepsilon^{2}
\end{eqnarray}
if we disregard the terms associated to the relative motion.

\section{The calculation of the butterfly energy spectrum}

The total wave function of the zeroth-order Hamiltonian, after symmetrization,
is given by 
\begin{equation}
|IMKn_{\varepsilon}\rangle =
 \left [\frac{2I+1}{16\pi^{2}(1+\delta_{K0})}\right]^{1\over 2}
\left (D_{MK}^{I\dagger}(\Omega)+
(-)^{I}D_{M-K}^{I\dagger}(\Omega)\right)
\chi_{K,n_{\varepsilon}}(\varepsilon)
\end{equation}
where \cite{eg75}
\begin{equation}
\chi_{K,n_{\varepsilon}}=
\frac{\{\lambda^{l_{K}+{3\over 2}}\Gamma(l_{K}+{3\over 2}+n_{\varepsilon})\}}
{(n_{\varepsilon}!)^{1/2}\Gamma(l_{K}+{3\over 2})}
|\varepsilon|^{1/2}\varepsilon^{K/2}e^{-{1\over 2}\lambda\varepsilon^{2}}
{~}_{1}F_{1}(-n_{\varepsilon},l_{K}+{3\over 2};\lambda\varepsilon^{2})
\end{equation}
is the wave function describing the butterfly harmonic vibrations.
The energy spectrum is then
\begin{equation}
E_{IKn_{\varepsilon}}=(l_{K}+2n_{\varepsilon}+{3\over 2})
\hbar\omega_{\varepsilon} + (I(I+1)-K^{2})\frac{\hbar^{2}}{2\mu r_{0}^{2}}
\end{equation}
with $n_{\varepsilon}$=0,1,2,\ldots; $K$=0,2,4,\ldots; $I$=0,2,4,\ldots if 
$K$=0 and $I$= $K$, $K$+1, $K$+2, \ldots if $K\neq$0.
The butterfly oscillation frequency is
$$
\omega_{\varepsilon} = \sqrt{\frac{C_{\varepsilon}}{2{\cal J}_{12}}}
$$
with ${\cal J}_{12}=3[B_{1}\beta_{01}^{2}+
{R_{1}^2\over R_{2}^2}B_{2}\beta_{02}^{2}]$ being the inertia parameter 
of the nuclear molecule.
We notice that this model works only for even-even splittings 
of $^{252}$Cf. In order to extend the calculations to the odd-odd 
splittings the particle-core coupling must pe appropriately included in 
the Hamiltionian. We will address this question in a forthcoming paper  
In Table I we list the zero-energy $\hbar\omega_{\varepsilon}$ for nuclear 
molecule configurations corresponding to even-even splitings of $^{252}$Cf. 
It is important to substantiate once again the importance of
fragments deformation in the excitation of these states. When both fragments
are well deformed the first excited state
($\approx 2\hbar\omega_{\varepsilon}$) is lying at low energy. If one 
of the emitted fragments is nearly spherical the location is shifted up 
in energy up to 10 times or more. Therefore it is expected to observe these 
states for binary cold fission channels with fragments having large 
deformations. 
On the other hand, resuming the calculations from Table I, but 
this time with the same values for the deformations of the light and heavy 
fragments, we arrive at the conclusion that the energy of these vibrations 
depends only sligthly on the mass assymetry.   

\begin{table}
\caption{The value of butterfly zero-energy $\hbar\omega_{\varepsilon}$ 
for different even-even splittings of $^{252}$Cf} 
\begin{indented}
\item[]
\begin{tabular}{@{}ccccc}
\br
Light fragment & $\varepsilon_{L}$ & Heavy fragment & 
$\varepsilon_{H}$ & $\hbar\omega_{\varepsilon}$(KeV)\\
\br
$^{94}$Kr&0.270&$^{158}$Sm&0.300&517.88\\
\mr
$^{94}$Sr&0.215&$^{158}$Nd&0.295&636.42\\
\mr
$^{96}$Sr&0.240&$^{156}$Nd&0.320&607.68\\
\mr
$^{98}$Sr&0.330&$^{154}$Nd&0.230&609.36\\
\mr
$^{98}$Zr&0.255&$^{154}$Ce&0.283&603.70\\
\mr
$^{100}$Sr&0.350&$^{152}$Nd&0.250&509.43\\
\mr
$^{100}$Zr&0.273&$^{152}$Ce&0.267&623.05\\
\mr
$^{102}$Zr&0.340&$^{150}$Ce&0.200&767.60\\
\mr
$^{104}$Mo&0.315&$^{148}$Ba&0.245&597.80\\
\mr
$^{104}$Zr&0.300&$^{140}$Ce&0.210&767.60\\
\mr
$^{106}$Mo&0.245&$^{146}$Ba&0.200&1074.16\\
\mr
$^{106}$Zr&0.330&$^{146}$Ce&0.170&815.05\\
\mr
$^{108}$Mo&0.245&$^{144}$Ba&0.198&1088.19\\
\mr
$^{108}$Ru&0.300&$^{144}$Xe&0.200&805.73\\
\mr
$^{110}$Mo&0.240&$^{142}$Ba&0.150&1421.21\\
\mr
$^{112}$Ru&0.220&$^{140}$Xe&0.120&2051.47\\
\mr
$^{114}$Ru&0.180&$^{138}$Xe&0.060&3664.41\\
\mr
$^{114}$Pd&0.150&$^{138}$Te&0.140&2863.09\\
\mr
$^{116}$Pd&0.140&$^{136}$Te&0.060&4956.01\\
\mr
$^{116}$Ru&0.150&$^{136}$Xe&-0.020&4839.04\\
\mr
$^{118}$Pd&0.100&$^{134}$Te&0.000&7736.05\\
\mr
$^{120}$Cd&0.030&$^{132}$Sn&-0.010&14497.94\\
\mr
$^{120}$Pd&0.110&$^{132}$Te&-0.005&4076.04\\
\mr
$^{122}$Pd&0.008&$^{108}$Te&0.000&9213.6.1\\
\br
\end{tabular}
\end{indented}
\end{table}

\section{Concluding remarks}

The recent measurement of the double fine structure in the cold fission 
of $^{252}$Cf opened the interest to search for possible molecular states  
in the fragmentation process where two or more nuclei are involved. 
The believe on the existence of such collective excitations is supported 
by the fact that cold fission is just a natural extension of cluster 
radioactivity to heavy nuclei. 
In this paper we predicted a molecular vibrational mode for the cold 
fission of $^{252}$Cf in two fragments with axial symmetric deformation. 
Geometrically this mode may be pictured as in-phase rotational vibrations 
of the two fragments with respect to an axis perpendicular to their 
symmetry axis which passes through the touching point. 
The magnitude of the butterfly quantum energy depends strongly on the   
ground state deformations of both fragments.
Concerning the possibility to observe $\gamma$ transitions
from butterfly-type molecular states we make the observation that  
such a project is feasible. The large bulk of experimental data 
provided by the Gamma sphere facility from ORNL on the $\gamma$ 
decay transitions from low-lying states of individual fragments 
will turn out to be the corner stone in the search of similar 
transitions but of nuclear molecular nature.
As for the future we intend to extend these calculations in order to 
encompass also the odd-odd splittings and to study the influence of 
higher multipolarity deformations like octupole and hexadecupole on 
the molecular vibrational spectra.

\Figures
\Figure{Butterfly-type oscillations of a nuclear molecule}


\section*{References}
\begin{thebibliography}{99}
\bibitem{hkb93}Knitter H-H, Hambsch F J and Budtz-Jorgensen C 
1993 \NP {\bf A~554} 209 
\bibitem{sfc96}S\u andulescu A, Florescu A, C\^{a}rstoiu F, Greiner W, 
Hamiltion J H, Ramayya A V and Babu B R S 1996 \PR {\bf C~54} 258
\bibitem{tak94}Ter-Akopian G M {\em et al} 1994 \PRL 
{\bf 73} 1477
\bibitem{gps95} Greiner W, Park Jae Y, Scheid W 1995 {\em Nuclear 
Molecules} World Scientific, Singapore
\bibitem{spg80} S\u andulescu, Poenaru D, Greiner W 1980 {\em Sov. J.
Part. Nucl.} {\bf 11} 528
\bibitem{iac81} Iachello F 1981 \PR {\bf C~23} 2778
\bibitem{hg84}Hess P O and Greiner W 1984 \NC 
{\bf 83A} 76 
\bibitem{eg75}Eisenberg J M and Greiner W 1975 {\em Nuclear Theory I}
North-Holland, Amsterdam
\end{thebibliography}
\end{document}